\documentclass[preprintnumbers,amsmath,amssymb,twocolumn]{revtex4}
\usepackage{mathrsfs}
\usepackage{graphicx}
\usepackage[a4paper,height=22cm,width=18cm]{geometry}
\usepackage{dcolumn}
\usepackage{bm}
\usepackage[dvips]{color}

\allowdisplaybreaks[4]

\usepackage{latexsym}
\usepackage{palatino}
\usepackage{makeidx,ifthen}
\usepackage{amsmath}
\usepackage{amssymb}
\usepackage{dcolumn}
\usepackage{bm}
\usepackage[dvips]{color}

\newcommand{\beq}{\begin{equation}}
\newcommand{\eeq}{\end{equation}}
\newcommand{\bq}{\begin{equation}}
\newcommand{\eq}{\end{equation}}
\newcommand{\ba}{\begin{array}}
\newcommand{\ea}{\end{array}}
\newcommand{\bea}{\begin{eqnarray}}
\newcommand{\eea}{\end{eqnarray}}
\def\bc{\begin{center}}
\def\ec{\end{center}}
\def\bnum{\begin{enumerate} }
\def\enum{\end{enumerate}}

\def\[{\left[}
\def\]{\right]}
\def\({\left(}
\def\){\right)}

\def\>{\rightarrow}

\def\ETslash{\not{\hbox{\kern-4pt $E_{T}$}}}

\def\Dslash{\not{\hbox{\kern-4pt $D$}}}
\def\pslash{\not{\hbox{\kern-4pt $p$}}}
\def\qslash{\not{\hbox{\kern-4pt $q$}}}
\def\lv{\not{\hbox{\kern-4pt $L$}}}
\def\lsim{\mathrel{\raise.3ex\hbox{$<$\kern-.75em\lower1ex\hbox{$\sim$}}}}
\def\gsim{\mathrel{\raise.3ex\hbox{$>$\kern-.75em\lower1ex\hbox{$\sim$}}}}
\def\ifmath#1{\relax\ifmmode #1\else $#1$\fi}

\evensidemargin=\oddsidemargin

\begin{document}
\begin{flushleft}
TUHEP-TH-10172
\end{flushleft}

\null\vspace{-1cm}
\title{Anomalous gauge couplings of the Higgs boson at the LHC II: Study of additional backgrounds in semileptonic mode of WW scatterings}

\author{ Yong-Hui Qi \footnote{Email:\,qiyh05@mails.tsinghua.edu.cn,} and Yu-Ping Kuang \footnote{Email:\,ypkuang@mail.tsinghua.edu.cn}}

\null\vspace{0.2cm}
\affiliation{Center for High Energy Physics and Department of Physics, Tsinghua University, Beijing 100084, China}

\begin{abstract}
In addition to a previous paper \cite{QKLZ09}, we study the backgrounds in which leptons coming from jet hadronization and from $\tau$ decay to the process
$pp\to W^+W^\pm j^f_1 j^f_2\to l^+\nu^{}_lj^{}_1j^{}_2j^f_1j^f_2$ for measuring the anomalous gauge couplings of the Higgs boson at the LHC.
We show that these kinds of backgrounds can be effectively suppressed by the cuts already imposed in Ref.\,\cite{QKLZ09}, so that all the conclusions in Ref.\,\cite{QKLZ09} will not be affected by these kinds of backgrounds.
\\\\
\null\noindent
\end{abstract}

\maketitle

Searching for the Higgs boson is of first priority in LHC experiments. Once a Higgs candidate is found, we then need to know whether it is the standard model (SM) Higgs boson or a Higgs boson in new physics beyond the SM. So far we do not know what model of new physics will actually reflect the property of the nature. A model-independent way of measuring the anomalous couplings of the Higgs boson will provide a no-lose study, i.e., if we find nonvanishing anomalous couplings (deviation from the SM couplings), we can conclude that it is a new physics effect.
In a previous paper \cite{QKLZ09}, we gave a hadron level study of a model-independent test of the anomalous gauge couplings of the Higgs boson at the 14 TeV LHC via the semileptonic mode of weak-boson scatterings $pp\to W^+W^\pm j^f_1 j^f_2\to l^+\nu^{}_lj^{}_1j^{}_2j^f_1j^f_2$, and the conclusion is that, with certain kinematical cuts imposed, the measurement can start for an integrated luminosity of 50 fb$^{-1}$. For higher integrated luminosity, say 100 fb$^{-1}$, the two main anomalous coupling constants $f^{}_{WW}/\Lambda^2$ and $f^{}_W/\Lambda^2$ can be determined separately which may provide a clue for figuring out the underlying theory of new physics.

 For convenience, we summarize the cuts imposed in Ref.\,\cite{QKLZ09} as follows.\\
\begin{description}
\item{\bf i}, charged Lepton transverse momentum cut
\begin{eqnarray}                         
&&p^{}_T(l^+)>200~{\rm GeV} \label{p_T(l^+)}.
\label{p_T(l)}
\end{eqnarray}
\item{\bf ii}, tagging two forward jets cut
\begin{eqnarray}                                
&& p^{}_T(j^f_i)> 20~{\rm GeV},~~~~
E(j^f_i)>300~{\rm GeV},\nonumber\\
&&2.0<|\eta(j^f_i)|<4.5,~~i=1,2, \eta(j^f_1)\eta(j^f_2)<0. \label{forward jet cuts}
\end{eqnarray}
\item{\bf iii}, hadronic decay of $W^\pm$ (Single jet $J$ with the largest $p_{T}$ among the reconstructed jets via $k_{T}$ algorithm): transverse momentum cut
\begin{eqnarray}                                     
 p^{}_T(J)>200~{\rm GeV},~~~~~~\eta(J)\eta(l^+)<0,
 \label{p_T(J)}
\end{eqnarray}
\item{\bf iv}, hadronic decay of $W^\pm$: invariant mass cuts
\begin{eqnarray}                                      
 65~{\rm GeV}<M_J<95~{\rm GeV}, \label{M_J}
\end{eqnarray}
\item{\bf v}, cuts reconstructing the electroweak scale
\begin{eqnarray}                                      
1.6<\log(p^{}_T\sqrt{y})<2.0 \label{log cut}
\end{eqnarray}
where $y\equiv \displaystyle\frac{d_{12}}{d_{cut}}$, $d_{12}=2{\rm min}(E_1^2,E_2^2)(1-\cos\theta_{12})$ [$E^{}_1,~E^{}_2,~ \theta^{}_{12}$ are the energies and relative angle of the first two sub-jets before emerging into the single jet $J$, clustered by the $k_T^{}$ algorithm, respectively (cf. Ref.\,\cite{Catani})], and $d_{cut}$ is the total energy by summing over all the reconstructed jets in each event.
\item{\bf vi}, top quark veto in order to suppressing top quark background
\begin{eqnarray}                                     
130~{\rm GeV}< M_{Jj} <240~{\rm GeV}. \label{top veto}
\end{eqnarray}
\item{\bf vii}, $p_{T}$ balance cuts for reconstructing $\l^{+} \ETslash jj j_{f}j_{f}$
\begin{eqnarray}                       
\sum_i p^{i}_T <\pm15~{\rm GeV}, \label{p_T balance}.
\end{eqnarray}
\item{\bf viii}, {\it minijet veto}: vetoing the events containing jets other than the signal jet $J$ from
$W^\pm$ decay [satisfying (\ref{p_T(J)}) and (\ref{M_J})] in the central rapidity region, $|\eta|<2$.
\end{description}

In Ref.\,\cite{QKLZ09}, the leptons, $l^+=e^+,\mu^+$, in the final state of the background processes are considered only from direct $W^+$ decays. However, in practical data analysis, there are other possibilities that the leptons may come from other sources mimicking the signal leptons, such as from the jet(s) hadronization ($l^+$ comes from the decays of pions, $\eta$, $J/\psi$, etc. in the hadronization of jets) and from $\tau$ decay, and thus become additional backgrounds to the signal process. In this short paper, we study these kinds of additional backgrounds and their suppressions for making the whole study more realistic.

The first kind serves as a background only at the hadron level when the hadronization of the jet is taken into account since, at the parton level, the final state particles are not the same as those in the signal process. For this kind of background, we shall study the following processes:
\beq                                    
pp\to Wjjj\to jjj_{f}j_{f}l^{+}\ETslash,
\label{W3j}
\eeq
\beq                                    
pp\to Zjjj\to jjj_{f}j_{f}l^{+}\ETslash,
\label{Z3j}
\eeq
\beq                                     
pp\to jjjj\to Jj_{f}j_{f}l^{+}\ETslash,
\label{4j}
\eeq
\beq                                     
pp\to jjjjj\to jjj_{f}j_{f}l^{+}\ETslash,
\label{5j}
\eeq
in which the lepton $l^+$ comes from the hadronization of jets, and the two jets $j^{}_f j^{}_f$ are chosen to satisfy the forward-jet cuts (\ref{forward jet cuts}). The remained two jets $jj$ in (\ref{W3j}), (\ref{Z3j}) and (\ref{5j}) correspond to the jets in $W(Z)\to jj$, and will be required to behave as a ``single'' energetic jet $J$ satisfying the cuts (\ref{p_T(J)}) and (\ref{M_J}).

For the second kind of background, we shall study the following backgrounds in addition to those studied in Ref.\,\cite{QKLZ09}:
\bea                                    
pp\to Zj\to \tau^+\tau^-j\to l^+Jj^{}_fj^{}_f+\ETslash,
\label{Z+j}
\eea
where $l^+$ is from $\tau^+$ decay; and
\bea                                    
pp\to Zjj\to \tau^+\tau^-jj\to l^+jjj^{}_fj^{}_f+\ETslash,
\label{Z+2j}
\eea
where $l^+$ is from $\tau^+$ decay, and the two jets $jj$ correspond to the jets in $W(Z)\to jj$ in the signal process, which should be required to behave as a ``single'' energetic jet $J$ satisfying the cuts (\ref{p_T(J)}) and (\ref{M_J}).

\bea                                    
pp\to Wjjj\to \tau^+\nu^{}_\tau jjj\to l^+jjj^{}_fj^{}_f\ETslash,
\label{W+3j}
\eea
where $l^+$ is from $\tau^+$ decay, and the two jets, $jj$, correspond to the jets from $W$ decay but actually not necessarily behave as an energetic single-jet $J$; and
\bea                                    
pp\to Zjjj\to \tau^+\tau^-jjj\to l^+jjj^{}_fj^{}_f\ETslash,
\label{Z+3j}
\eea
where $l^+$ and $jj$ are similar to those in (\ref{W+3j}).

Now we study the above additional backgrounds at the hadron level under the cuts summarized in (\ref{p_T(l)})$\--$(\ref{p_T balance}) and {\it minijet veto} taking account of both the initial and final state partons radiations, following the parton showers and hadronization of the final state jets by using PYTHIA \cite{PYTHIA} and the $k^{}_T$ algorithm \cite{Catani} with the package ALPGEN \cite{ALPGEN}.

\begin{widetext}

\begin{figure}[h]
\includegraphics[scale=0.32]{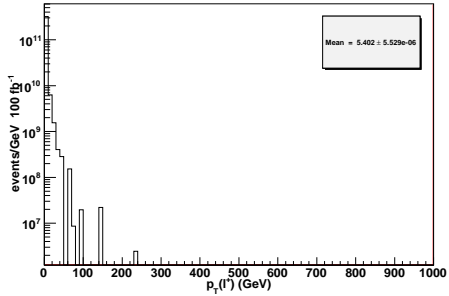}\quad\includegraphics[scale=0.32]{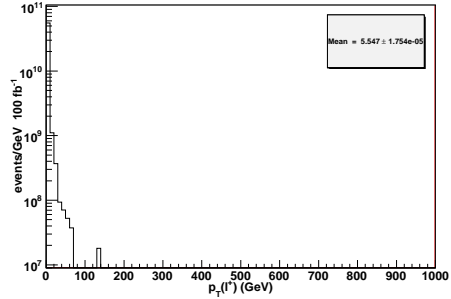}\\
\null\vspace{-0.8cm}
\null\hspace{0cm}(a)\hspace{6.5cm}(b)
\caption{$p^{}_T(l^+)$ distribution in the process (\ref{4j}) and (\ref{5j}) for an integrated luminosity of 100 fb$^{-1}$: (a) in the process (\ref{4j}), and (b) in the process (\ref{5j}) [those in the processes (\ref{W3j}) and (\ref{Z3j}) are similar to (b)].}
\label{5j:p_T}
\end{figure}
\end{widetext}
In the backgrounds (\ref{W3j})$\--$(\ref{5j}), we naturally estimate that the transverse momenta of such leptons are of the order of the hadronization scale which are much smaller than the transverse momentum of the signal lepton from $W^+$ decays. Thus we expect that the imposed cut (\ref{p_T(l)})
can effectively suppress these backgrounds. After the hadronic level calculation using PYTHIA \cite{PYTHIA} and the $k^{}_T$ algorithm \cite{Catani} with the package ALPGEN \cite{ALPGEN}, 
we obtain the $p^{}_T(l^+)$-distributions in the processes (\ref{W3j})$\--$(\ref{5j}). The distributions in (\ref{W3j})$\--$(\ref{5j}) are all similar. As an example, we show the $p^{}_T(l^+)$-distribution of the process (\ref{4j}) and (\ref{5j}) in FIG.\,\ref{5j:p_T}. We see that the distribution is indeed mainly in the low $p^{}_T(l^+)$ region, but there are still some possibilities of being in higher $p^{}_T$ regions. We shall show that, with other cuts in (\ref{p_T(l)})$\--$(\ref{p_T balance}), these backgrounds are suppressed to be negligibly small.

\begin{widetext}

\begin{figure}[h]
\bc
\null\vspace{-0.5cm}
\includegraphics[scale=0.32]{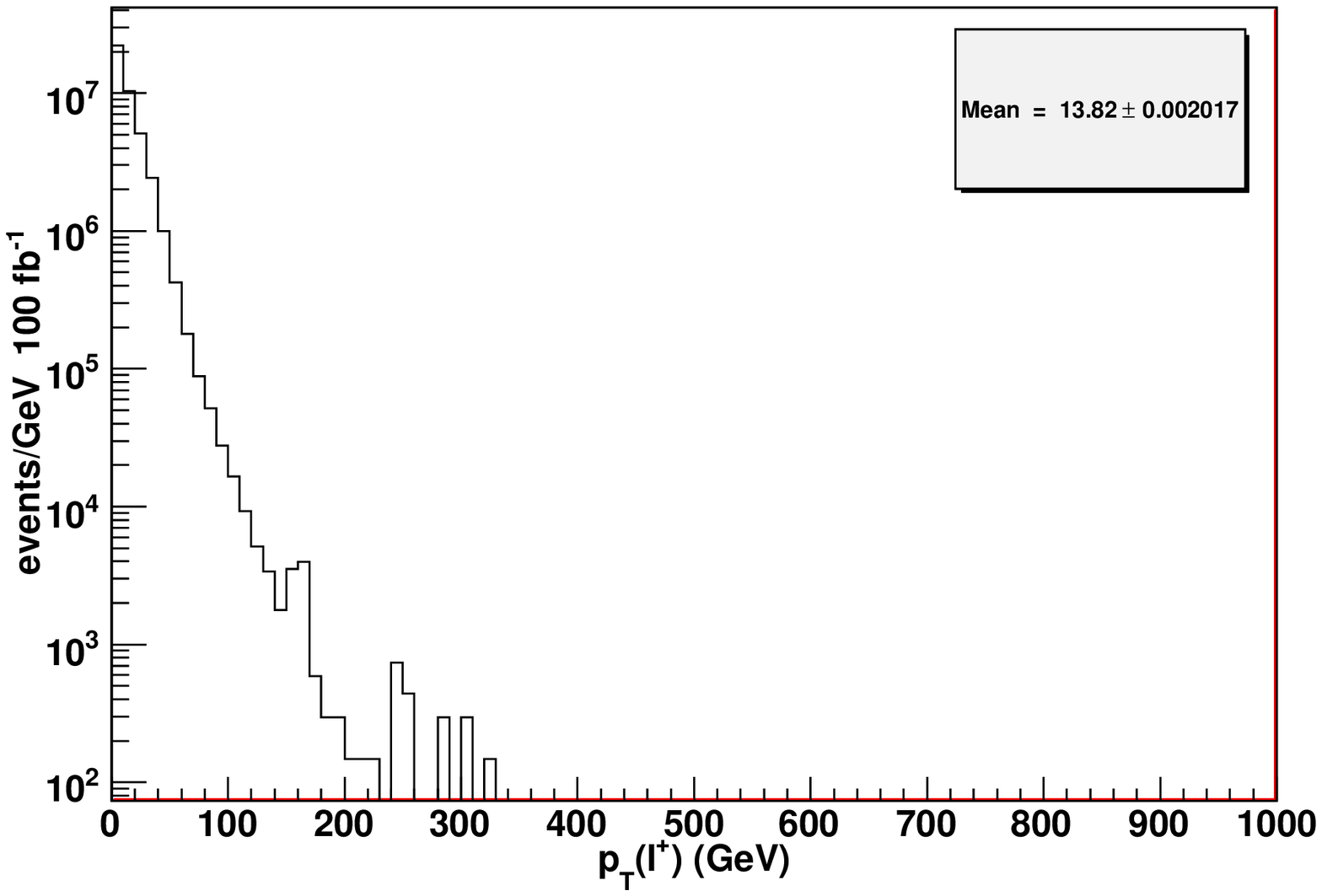}\quad\includegraphics[scale=0.32]{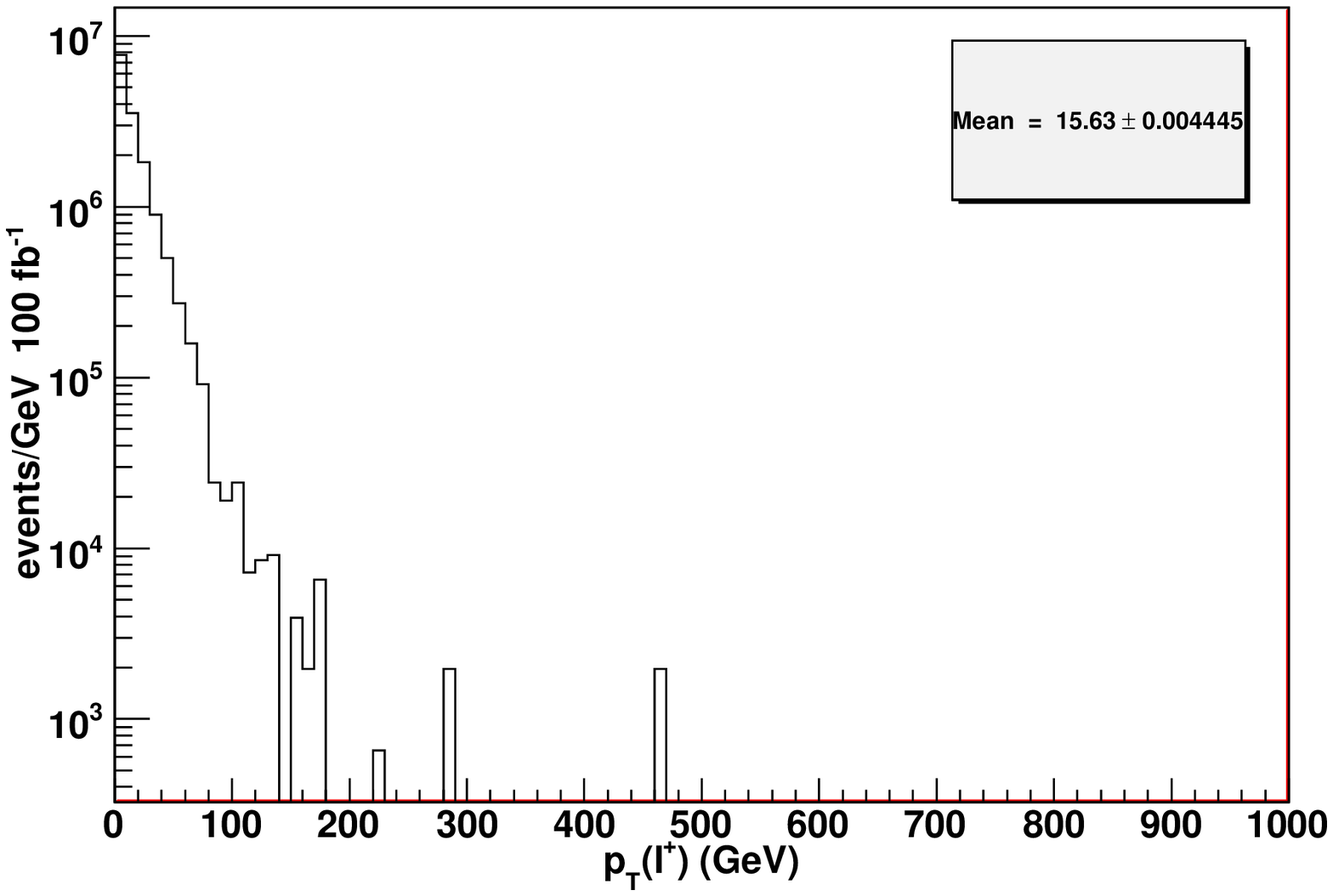}\\
\null\vspace{-0.4cm}\hspace{0cm}(a)\hspace{6.5cm}(b)
\ec
\null\vspace{-0.5cm}
\bc
\null\vspace{-0.65cm}
\includegraphics[scale=0.32]{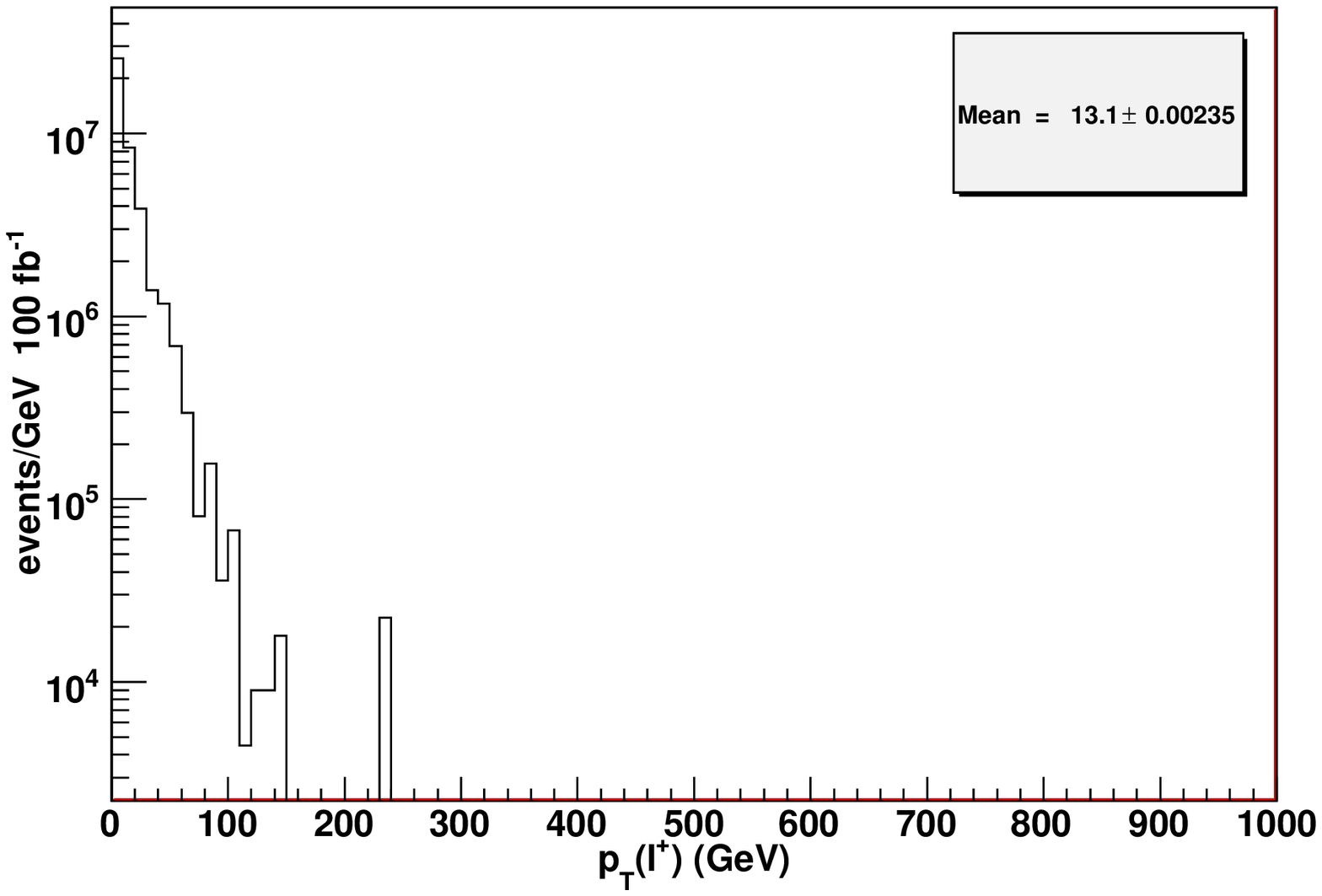}\quad\includegraphics[scale=0.32]{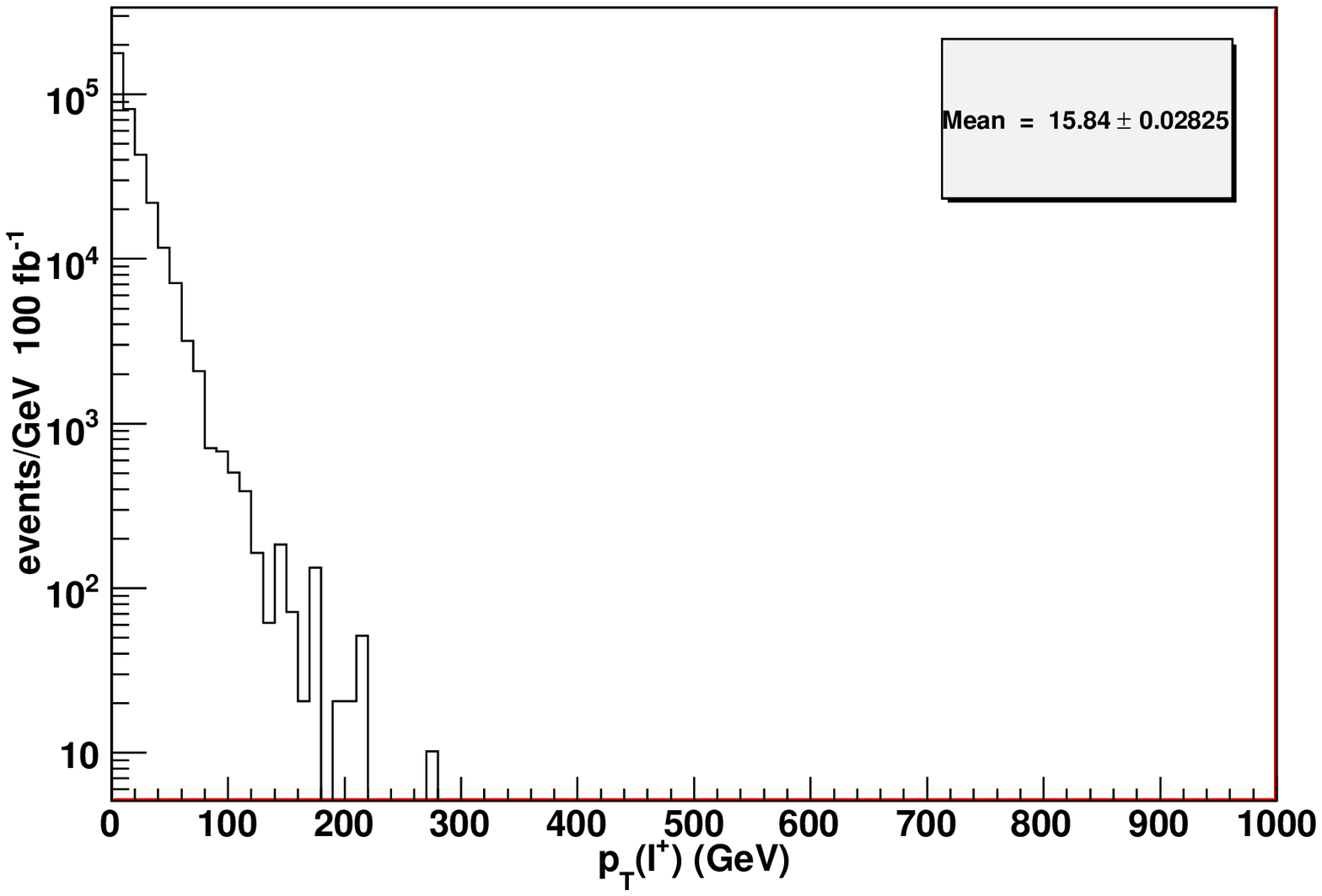}\\
\null\vspace{-0.4cm}\hspace{0cm}(c)\hspace{6.5cm}(d)
\ec
\null\vspace{-0.85cm}
 \caption{\label{tau:p_T} $p^{}_T(l^+)$ distributions in the background processes (\ref{Z+j})$\--$(\ref{Z+3j}).
 with an integrated luminosity of 100 fb$^{-1}$: (a) $p^{}_T(l^+)$ distribution in (\ref{Z+j}); (a) $p^{}_T(l^+)$ distribution in (\ref{Z+j}); (b) $p^{}_T(l^+)$ distribution in (\ref{Z+2j}); (c) $p^{}_T(l^+)$ distribution in (\ref{W+3j}); and
  (d) $p^{}_T(l^+)$ distribution in (\ref{Z+3j}).
   }
\end{figure}
\end{widetext}

In FIG.\,\ref{tau:p_T}, we plot the corresponding $p^{}_T(l^+)$ distributions in the background processes (\ref{Z+j})$\--$(\ref{Z+3j}). We see that, in these three processes, there are $p^{}_T(l^+)$ distributed above 200 GeV. Therefore other cuts in (\ref{p_T(l)})$\--$(\ref{p_T balance}) and {\it minijet veto} are also needed for suppressing these backgrounds.

To have a complete list of the efficiency of the imposed cuts in suppressing all the signals and backgrounds studied in the present paper and in Ref.\,\cite{QKLZ09}, we illustrate the complete cut efficiency in TABLE\,\ref{efficiency} by taking the case of $m^{}_H=115$ GeV with $f^{}_W/\Lambda^2=4.0$ TeV$^{-2}$ as an example. In TABLE\,\ref{efficiency}, the column labeled by ``particle ID'' means that we take only the basic cuts for identifying particles taken in Ref.\,\cite{QKLZ09} according to the detecting limits of the detector (ATLAS or CMS) without imposing the kinematic cuts {\bf i}$\--${\bf viii}.

We see from TABLE\,\ref{efficiency} that all the additional backgrounds studied in this paper (the first eight rows) are made negligibly small by imposing the first four cuts, {\bf i}$\--${\bf iv} [Eqs,\,(\ref{p_T(l)})$\--$(\ref{M_J})]. Therefore, {\it with the cuts (\ref{p_T(l)})$\--$(\ref{p_T balance}) and {\it minijet veto} already imposed in Ref.\,\cite{QKLZ09} (without adding any new cuts), all the conclusions stated in Ref.\,\cite{QKLZ09} are not affected by taking account of the additional backgrounds studied in this paper}. So we state our final conclusion as follows. {\it With the kinematic cuts imposed in Ref.\,\cite{QKLZ09}, a model-independent test of the anomalous gauge couplings of the Higgs boson at the 14 TeV LHC via the semileptonic mode of weak-boson scatterings $pp\to W^+W^\pm j^f_1 j^f_2\to l^+\nu^{}_lj^{}_1j^{}_2j^f_1j^f_2$ can be obtained by imposing the kinematic cuts (\ref{p_T(l)})$\--$(\ref{p_T balance}) and {\it minijet veto}. The experimental measurement can start for an integrated luminosity of 50 fb$^{-1}$. For higher integrated luminosity, say 100 fb$^{-1}$, the two main anomalous coupling constants $f^{}_{WW}/\Lambda^2$ and $f^{}_W/\Lambda^2$ can be quantitatively determined separately to a certain accuracy \cite{QKLZ09} which may provide a clue for figuring out the underlying theory of new physics}.
\begin{widetext}

\begin{table}[h]
 \caption{ Cut efficiency of the cross sections (in fb) for the above additional backgrounds (\ref{W3j})$\--$(\ref{Z+3j}), the signal plus irreducible backgrounds (IB), and the reducible backgrounds considered in Ref.\,\cite{QKLZ09} with the Higgs boson mass $m_{H}=115$ GeV, and the  anomalous coupling $f_{W}/\Lambda^2=4.0~ \text{TeV}^{-2}$ (with other anomalous
 couplings vanishing) as an example.
 All processes are studied up to the hadron level. The column labeled by ``particle ID'' means that we take only the basic cuts for identifying particles (cf. Ref.\,\cite{QKLZ09}) according to the detecting limits of the detector (ATLAS or CMS) without imposing the kinematic cuts {\bf i}$\--${\bf viii}.}
 \null\vspace{-0.2cm}
 \tabcolsep 1.0pt
\begin{tabular}{ccccccccc}
\hline  \hline
Cuts  &  particle ID & cut={\bf i} [(\ref{p_T(l^+)})] & cut-{\bf ii} [(\ref{forward jet cuts})] &cuts {\bf iii, iv} [(\ref{p_T(J)}), (\ref{M_J})]& cut-{\bf v} [(\ref{log cut})] & cut-{\bf vi} [(\ref{top veto})]
&cut-{\bf vii} [(\ref{p_T balance})] & cut-{\bf viii}: Minijet veto \\
\hline
W+3j [(8)]&2525616&0&-&-&-&-&-&-\\
Z+3j [(9)]&511353&0&-&-&-&-&-&-\\
$4j$ [(10)]&3299949610&24566.5&0&-&-&-&-&-\\
$5j$ [(11)]&572442776&0&-&-&-&-&-&-\\
Z+j [(12)]&417519&23.57&1.47&0&-&-&-&-\\
Z+2j [(13)]&151189&45.7752&0&-&-&-&-&-\\
W+3j [(14)]&420936&224.356&44.8711&0&-&-&-&-\\
Z+3j [(15)]&3525.88&0.818498&0.204625&0.102312&0&-&-&-\\

\hline
signal with IB in Ref.\,\cite{QKLZ09}&210.66&34.55&11.29&7.01&2.42&2.39&2.28&2.28\\
IB ($f_W=0$) in Ref.\,\cite{QKLZ09}&338.82&36.08&9.44&4.12&1.29&1.27&1.26&1.26\\
WZ+2-jet in Ref.\,\cite{QKLZ09}&1431.67&36.93&2.40&0.12&2.7$\times{10^{-2}}$&2.3$\times{10^{-2}}$&5$\times{10^{-4}}$&-\\
W+3-jet in Ref.\,\cite{QKLZ09}&2908923&9630.86&104.25&0.10&6.1$\times{10^{-3}}$&4.7$\times{10^{-3}}$&2$\times{10^{-4}}$&-\\
$t\bar{t}$ in Ref.\,\cite{QKLZ09}&407776.84&2586.47&61.77&1.09&0.09&0.06&2$\times{10^{-3}}$&-\\
\hline
\hline
\end{tabular}
\label{efficiency}
\end{table}

\end{widetext}

We would like to thank Ming-Shui Chen for valuable discussions. This work is supported by National Natural Science Foundation of China under Grant Nos. 10635030 and 10875064.

\newpage
\bibliography{000}

\begin{thebibliography}{0}

\bibitem{QKLZ09}
  Y.~H.~Qi, Y.~P.~Kuang, B.~J.~Liu and B.~Zhang,
    Phys.\ Rev.\  D {\bf 79}, 055010 (2009)
  [arXiv:0811.3099 [hep-ph]].

\bibitem{PYTHIA}
T. Sj\"{o}strand, P. Ed¡äen, C. Friberg, L. L\"{o}nnblad, G.
Miu, S. Mrenna and E. Norrbin, Computer Physics Commun {\bf 135}, 238 (2001).

\bibitem{Catani}
S. Catani et al., Nucl. Phys. {\bf B 406}, 187 (1993); M.H.Seymour, Z. Phys. C {\bf 62}, 127 (1994).

\bibitem{ALPGEN}

  M.~L.~Mangano, M.~Moretti, F.~Piccinini, R.~Pittau and A.~D.~Polosa,
  JHEP {\bf 0307}, 001 (2003)
  [arXiv:hep-ph/0206293].


\end{thebibliography}

\end{document}